# Weak ferromagnetic order breaking the threefold rotational symmetry of the underlying kagomé lattice in CdCu$_3$(OH)$_6$(NO$_3$)$_2$·H$_2$O

Ryutaro Okuma*, Takeshi Yajima, Daisuke Nishio-Hamane, Tsuyoshi Okubo, and Zenji Hiroi

*Institute for Solid State Physics, University of Tokyo, Kashiwa, Chiba 277-8581, Japan*

Novel magnetic phases are expected to occur in highly frustrated spin systems. Here we study the structurally perfect kagomé antiferromagnet CdCu$_3$(OH)$_6$(NO$_3$)$_2$·H$_2$O by magnetization, magnetic torque, and heat capacity measurements using single crystals. An antiferromagnetic order accompanied by a small spontaneous magnetization that surprisingly is confined in the kagomé plane sets in at $T_N \sim 4$ K, well below the nearest-neighbor exchange interaction $J / k_B = 45$ K. This suggests that a unique "**q** = 0" type 120º spin structure with "negative" (downward) vector chirality, which breaks the underlying threefold rotational symmetry of the kagomé lattice and thus allows a spin canting within the plane, is exceptionally realized in this compound rather than a common one with "positive" (upward) vector chirality. The origin is discussed in terms of the Dzyaloshinskii–Moriya interaction.

## I. INTRODUCTION

Highly frustrated antiferromagnets with triangle-based lattices have been extensively studied both theoretically and experimentally, lured by a fascinating conjecture that the geometrical frustration enforces spins to disorder even at low temperatures and realizes exotic ground states defeating the conventional Néel order. In particular, strong frustration in kagomé antiferromagnets (KAFMs) with nearest-neighbor magnetic interactions $J$s in the kagomé net made of corner-sharing triangles has been focused on, because it should result in macroscopic degeneracy in the ground state of the classical Heisenberg model,[1-4] while, for the spin-1/2 Heisenberg model, one expects $Z_2$ gapped or gapless $U(1)$ Dirac quantum spin liquid states induced by large quantum fluctuations.[5,6] However, these exotic ground states are elusive and tend to be superseded by certain non-collinear long-range orders (LROs). Nevertheless, the KAFM is intriguing as it may exhibit an unconventional LRO with an emerging chiral degree of freedom on every triangle. The chirality is an important ingredient to recent condensed matter physics because it can generate topological spin textures such as skyrmions[7] and may couple with lattice, polarization and conduction electrons in various ways to generate novel phenomena.[8,9]

There are often additional interactions that may transform spin liquid states into LROs in the spin-1/2 KAFM, which are inter-plane couplings, further-neighbor interactions, exchange anisotropy, Dzyaloshinskii–Moriya (DM) interaction and so on. For example, the DM interaction prefers a coplanar **q** = 0 order [Figs. 1(a) and (b)] when its magnitude is larger than 10% of $J$.[10] On the other hand, ferromagnetic second-neighbor interactions favor the **q** = (1/3 1/3) order [Fig. 1(c)], and, further-neighbor interactions may stabilize more complex, noncoplanar spin structures with larger unit cells such as the twelve-sublattice order called the cuboc order.[11]

The three coplanar 120˚ structures illustrated in Fig. 1 are distinguished by the vector chirality $\kappa$ defined for each triangle by eq. (1):

$$\kappa = \frac{2}{3\sqrt{3}} (S_1 \times S_2 + S_2 \times S_3 + S_3 \times S_1). \quad (1)$$

Following the convention that the spins in the cross products appear rotating counterclockwise around the triangle, as shown in Fig. 1(a), the vector chirality points up and down normal to the kagomé plane in every triangle for Figs. 1(a) and (b), respectively, while is staggered in (c). Let us call the up and down $\kappa$ as 'positive' and 'negative', respectively, as widely used. Then, we call the three types of spin structures with different arrangements of $\kappa$ "positive" vector chirality (PVC), "negative" vector chirality (NVC), and staggered vector chirality (SVC) structures, respectively.

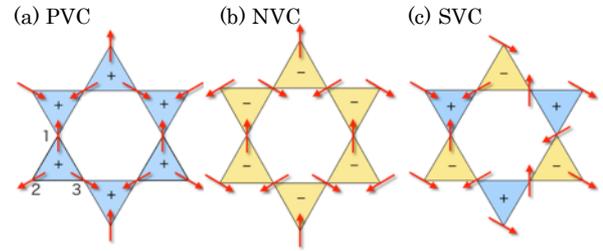

FIG. 1. Three coplanar 120º spin structures. A spin on each lattice point is represented by an arrow, and the direction of the vector chirality on each triangle is shown by '+' (up) or '–' (down). The two **q** = 0 type structures, "positive" vector-chirality (PVC) and "negative" vector-chirality (NVC) structures, are shown in (a) and (b), respectively, and the $\sqrt{3} \times \sqrt{3}$ structure with **q** = (1/3 1/3), a staggered vector-chirality (SVC) structure, in (c). The numbers around the lower-left triangle of (a) refer to those in eq. (1).

Among these three structures, the PVC order has been often observed in actual compounds. The $S = 5/2$ Fe jarosite KFe$_3$(OH)$_6$(SO$_4$)$_2$ exhibits a PVC order at 65 K,[12,13] while vesignieite, which is a structurally perfect or a slightly distorted $S = 1/2$ KAFM,[14,15] seems to show a PVC order at 9 K.[16] A PVC-type order with <100> anisotropy is also reported in quinternary oxalate compounds with a distorted kagomé lattice comprising Fe$^{2+}$.[17] On the other hand, a SVC order appears at 6 K in herbertsmithite, which is a structurally perfect $S = 1/2$ KAFM comprising $d(x^2 - y^2)$ orbitals of Cu$^{2+}$ ions[18] and seems to have a gapped spin liquid state,[19] when pressure above 2.5 GPa is applied.[20] In contrast, to our knowledge, there is no example of the NVC order for localized spin systems and thus has been rarely studied. The notable feature of NVC is that it loses all threefold rotation axes present in the underlying kagomé lattice unlike PVC or SVC. Therefore, it is intriguing to search for a KAFM that exhibits an NVC structure, which would provide us with a chance to study the property of this unusual magnetic order.

In the present study we focus on CdCu$_3$(OH)$_6$(NO$_3$)$_2$·H$_2$O[21] which is to be called Cd-kapellasite (CdK) for short. CdK is isostructural to kapellasite, ZnCu$_3$(OH)$_6$Cl$_2$:[22] compared to kapellasite, Cd$^{2+}$ and NO$_3^-$ ions are substituted for Zn$^{2+}$ and Cl$^-$ ions, respectively, and additional H$_2$O is intercalated in CdK (Fig. 2). CdK crystallizes in a trigonal structure with space group $P\bar{3}m1$ and lattice constants of $a = 6.522$ Å and $c =$

7.012 Å. There is a single Cu site which forms a $Cu(OH)_4(NO_3)_2$ octahedron heavily elongated toward the apical $NO_3^-$ ions by the Jahn–Teller effect. Thus, the $d(x^2 - y^2)$ orbital takes the highest $d$ level of the $Cu^{2+}$ ion and accommodates spin 1/2. The $Cu(OH)_4(NO_3)_2$ octahedra form a kagomé layer that contains an undistorted kagomé net of spin 1/2, as in herbertsmithite, with the Cd atom located at the center of the hexagon. The nitrate ions and water molecules separate the kagomé layers with an inter-plane distance as long as 7.012 Å, which is larger than 5.733 Å for kapellasite and is much larger than ~2 Å in the plane, indicating a good two dimensionality in magnetic interactions. There is one possible source of crystallographic disorder in CdK, which is associated with the configuration of the $NO_3^-$ unit.[21]

On the magnetic properties of CdK, the previous studies using polycrystalline samples found relatively large, antiferromagnetic Weiss temperatures of $\Theta_W = -114 \pm 27$ K[23] or $-62$ K.[24] A magnetic LRO is observed at $T_N = 4\sim5$ K, which is accompanied by a weak ferromagnetism. The large frustration factor of $\Theta_W/T_N > 15$ indicates a presence of significant magnetic frustration. In contrast, kapellasite and its Mg analogue, haydeeite, have nearly zero Weiss temperatures, suggesting that competing ferromagnetic and antiferromagnetic interactions largely compensate each other.[25] They exhibit no LRO above 2 K and a weak ferromagnetic order at 4 K, respectively. However, detailed magnetic properties, spin structures, and the origin of LRO are not known partly because of the lack of single crystalline samples.

In the present study, we have synthesized single crystals of CdK and performed magnetization and heat capacity measurements, which suggest that a NVC order is realized in CdK.

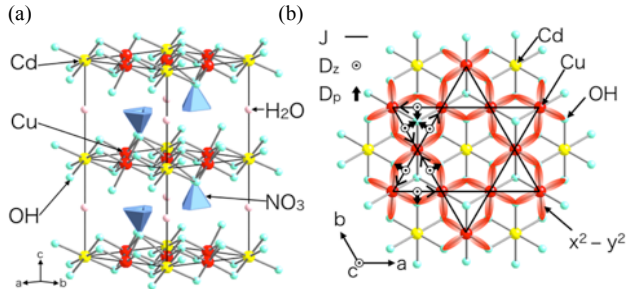

FIG. 2. Crystal structures of $CdCu_3(OH)_6(NO_3)_2 \cdot H_2O$ (CdK) (a) and the kagomé layer viewed along the [001] direction (b).[21] In (b), the arrangement of $d(x^2 - y^2)$ orbitals is depicted by the red lobes. The nearest-neighbor magnetic exchange interactions are shown by the thick lines. $D_z$ and $D_p$ in the pair of triangles represent the out-of-plane and in-plane components of the DM vector, respectively. The arrows on the pair of triangles represent the rotational direction in the cross products of the DM interactions, which is always counterclockwise.

## II. EXPERIMENTAL

Single crystals of CdK were synthesized by the hydrothermal transport method similar as used for the single crystal growth of herbertsmithite.[26] 5 g of $Cd(NO_3)_2 \cdot 4H_2O$ (98%, Sigma-Aldrich), 0.1 g of $Cu(OH)_2$ (90%, Wako Chemical) and 4 ml of deionized water were put into a quartz ampoule of 250 mm long and 12 mm in diameter, as shown in Fig. 3(a). Then, the ampoule was heated in a transparent furnace having a temperature gradient between 130 and 175 °C for a week. Light-blue precipitates first spreading over the ampoule were gradually transported to the cool zone, and many hexagonal blue crystals were finally obtained. The crystal of 0.5 mm both in edge and height shown in Fig. 3(b) was picked up and further examined.

A powder XRD pattern from crushed crystals was in good agreement with the calculated pattern, indicating that there is no impurity inclusion. Single crystal XRD results are consistent with the $P$–3$m$1 structure previously reported,[21] with slightly larger lattice constants of $a = 6.5449(7)$ Å and $c = 7.0328(9)$ Å. The chemical compositions determined by the inductively coupled plasma spectroscopy are 0.97(1) and 3.03(1) for Cd and Cu, respectively, which are close to the stoichiometric values. Magnetization was measured in a Magnetic Property Measurement System MPMS3 (Quantum Design), and both magnetic torque and heat capacity were measured in a Physical Property Measurement System PPMS (Quantum Design).

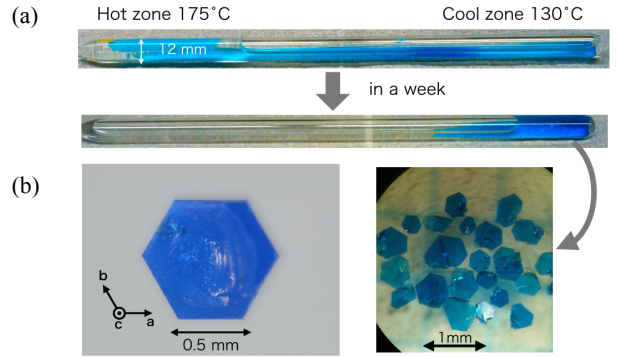

FIG. 3. (a) Crystal growth by the hydrothermal transport method and grown crystals. (b) Crystal of CdK used in the present experiments.

## III. RESULTS
### A. Magnetic susceptibility

Magnetic susceptibilities measured in magnetic fields of 1 T along the $a$ and $c$ axes exhibit Curie–Weiss-type temperature dependences at high temperatures above ~100 K and a broad hump at ~30 K, followed by a further increase at low temperature, as shown in Fig. 4. The hump is probably due to a short-range antiferromagnetic correlation, and the low-temperature enhancement may be associated with a weak ferromagnetic correlation mentioned later. Linear extrapolations of the reciprocal susceptibility give Weiss temperatures of approximately –60 K for both the directions, which agree with the previous value from a polycrystalline sample.[24] The small anisotropy is attributed to a difference in the Landé $g$ factor within the isotropic Heisenberg model. A simultaneous fit of the two datasets to the high-temperature series expansion for the $S = 1/2$ Heisenberg KAFM model[27] in the temperature range of 60–300 K [dotted lines in Fig. 4] yields $J / k_B = 45.44(3)$ K and $(g, \chi_0) = [2.2676(3), -3.62(7) \times 10^{-5}$ cm$^3$ Cu-mol$^{-1}$] and $[2.3330(6), -2.76(8) \times 10^{-5}$ cm$^3$ Cu-mol$^{-1}$] for the $a$ and $c$ directions, respectively. The temperature-independent term $\chi_0$ is from core diamagnetism (–$6 \times 10^{-5}$ cm$^3$ Cu-mol$^{-1}$) and Van-Vleck paramagnetism (not known). The $g$ values of CdK are along with typical values for Cu kagome minerals; $g = 2.1 \sim 2.2$ and $2.2 \sim 2.4$ for the parallel and perpendicular directions to the plane, respectively.[28]

We also fitted the magnetic susceptibility data to the high-temperature series expansion for the $J_1$–$J_2$–$J_d$ model employed to kapellasite;[29] an estimated set of $J$s for kapellasite are $J_1 = -12$ K, $J_2 = -4$ K and $J_d = 15.6$ K, which seems to generate short-range correlations toward a cuboc2 type order.[30] Because of many parameters, however, a reliable set of the $J$ values was not attained in our case. To be inferred safely from our fitting is that antiferromagnetic $J_2$ and ferromagnetic $J_d$ are smaller than half of dominant antiferromagnetic $J_1$. Thus, a simple nearest-neighbor model is a good starting point. The large differences in $J_1$ and $J_d$ between CdK and kapellasite may come from differences in the bridging anions (Cl$^-$ for kapellasite and NO$_3^-$ for CdK) and deeper $d$ levels of Cd than Zn.

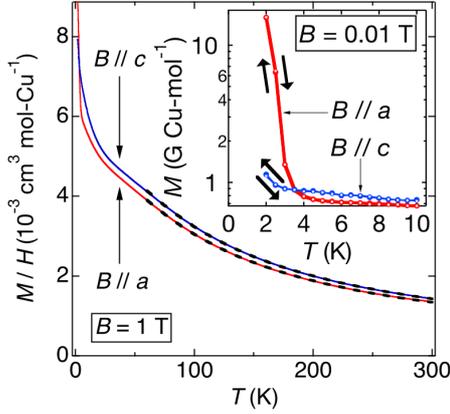

FIG. 4. Temperature dependence of magnetic susceptibilities from the one single crystal of CdK, the photograph of which is shown in Fig. 3(b). The measurements were carried out upon cooling in a magnetic field of 1 T applied along the $a$ or $c$ axis. The dashed lines on the data show fits to calculations by the high-temperature series expansion, which yields $J / k_B = 45.44(3)$ K. The inset shows magnetizations measured upon heating after zero-field cooling and then upon cooling below 10 K in a low field of 0.01 T.

## B. Magnetic order

A distinct anisotropy in magnetic susceptibility appears below ~4 K, as shown in the inset to Fig. 4: the magnetization suddenly increases for $B // a$, whereas remains less temperature-dependent for $B // c$. Corresponding to this anomaly in magnetization, heat capacity shows a broad peak at around 4 K, as shown in Fig. 5. These anomalies must indicate a long-range magnetic order at $T_N \sim 4$ K.

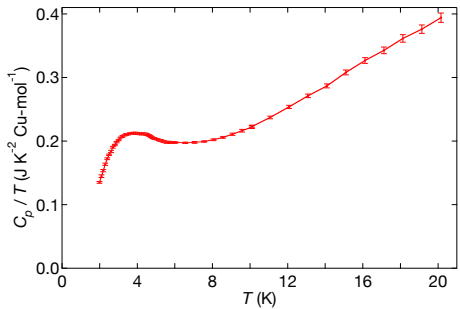

FIG. 5. Temperature dependence of heat capacity divided by temperature at zero magnetic field.

A possible reason why the heat capacity peak is so broad is an inhomogeneity associated with crystallographic disorder on the arrangement of the NO$_3^-$ unit.[21] Since one of the oxide ions of the NO$_3^-$ unit is located above or below the Cu triangle and mediates the nearest-neighbor superexchange coupling, this type of disorder may cause a spatial modulation in $J$. However, the disorder in CdK may not be strong enough to induce a spin glass freezing as no differences were observed between zero-field and field-cooled magnetic susceptibility curves at small $B$ [Fig. 4]. On the other hand, there can be an additional or alternative reason for the broadening at the magnetic transition, which will be addressed later.

## C. Weak ferromagnetism

Figure 6 shows magnetization processes at $T = 2$ K below $T_N$. Surprisingly, ferromagnetic behavior with a steep change near zero field is observed for $B // a$ and $a - b$, whereas is completely absent for $B // c$, implying that a ferromagnetic moment exists only within the kagomé plane! This is in distinct contrast to the out-of-plane weak ferromagnetism observed in the Cr jarosite.[31] Subtracting a linear component obtained by fitting $M$ in the range of 6–7 T from the $M$–$B$ curve for $B // a$ yields a small saturated magnetization of $7.93 \times 10^{-3}$ $\mu_B$ / Cu, just 0.8% of the expected magnetic moment for $S = 1/2$; almost the same value for $B // a - b$ indicates a weak anisotropy within the plane. A similar subtraction for $B // c$ gives a residual magnetization well fitted by the $S = 1/2$ Brillouin function with a nearly equal saturation. This means that the weak ferromagnetic moment lying in the plane is forced to align by the perpendicular field.

Since this weak ferromagnetism is observed only below $T_N$ and exhibits clear anisotropy, it should not originate from impurities, but must be parasitic to the antiferromagnetic order. Thus, it is plausible that a canted antiferromagnetic order is realized in CdK.

Alternative possibility is to ascribe such a small ferromagnetic moment to a magnetic domain wall (MDW). It was very recently found that uncompensated magnetic moments at MDWs in the three-dimensional all-in/all-out antiferromagnetic order of the pyrochlore oxide Cd$_2$Os$_2$O$_7$ give robust weak ferromagnetic moments.[32] Similar uncompensated moments at MDWs were observed in an oxalate compound with a PVC order on a distorted kagomé net.[17] In this case, however, the uncompensated moments behave as uncorrelated quasi-free moments at low temperature; it is unlikely that uncompensated moments at MDWs behave as correlated ferromagnetic moments in quasi-two-dimensional antiferromagnets. Therefore, we think it reasonable to assume that the weak ferromagnetism of CdK originates from a spin canting in the antiferromagnetic order.

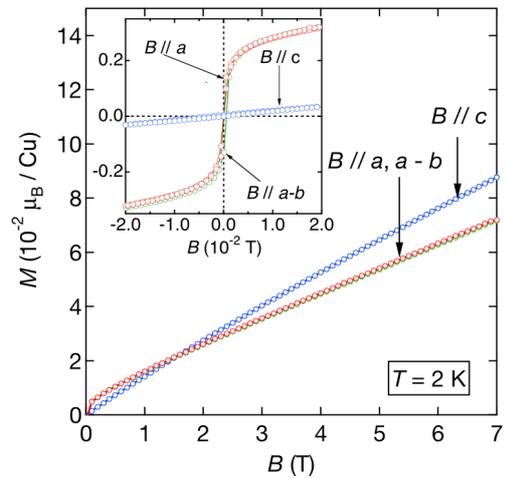

FIG. 6. Magnetization processes measured at 2 K. Blue, green and red marks are data for $B // c$, $a - b$, and $a$, respectively. The inset expands the small field range of $-0.02 \sim 0.02$ T.

## D. Magnetic anisotropy

The magnetic anisotropy is relatively small in the kagomé plane, as evidenced by the nearly equal M values for B // a and a − c in Fig. 6. However, there must be a tiny anisotropy reflecting the crystal symmetry. To investigate this, magnetic torque $\tau$ was measured by rotating the crystal around the c axis in a magnetic field of 10 T. Torque curves with clear 60° periodicity appear below $T_N$ and grow with decreasing temperature, as shown in Fig. 7.

The $\tau$ in the ab plane of a trigonal crystal system is given by eq. (2):

$$E(\phi) = K\cos 6\phi, \tau(\phi) = -\frac{\partial E}{\partial \phi} = K\sin 6\phi, \quad (2)$$

where $\phi$ is an angle from the easy or hard axis within the ab plane, and K is the magnetocrystalline anisotropy energy of the lowest order.[33] Since the experimental data are well reproduced by eq. (2), the trigonal crystal symmetry is actually preserved down to 2 K; no structural symmetry lowering takes place at low temperature. In addition, easy axes are determined at angles where the derivative of the torque curve becomes minimum: they are [100] and its equivalent directions, <100>. As a result, the local easy axes point to the center of the hexagon of the kagomé net, as depicted in the inset to Fig. 7. Note that the a − b direction corresponds to $\phi = 330°$, which is found to be a hard axis in the torque curves. Since the difference in M between the easy and hard axes is negligible, the anisotropy energy within the plane is quite small.

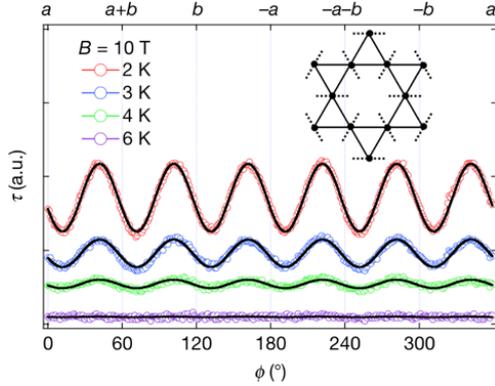

FIG. 7. Magnetic torque $\tau$ at various temperatures measured by rotating one crystal of CdK around the c axis in a magnetic field of 10 T. The rotation angle $\phi$ is set to zero at B // a. The solid line on each dataset is a fit to eq. (2). Easy axes appear with 60° interval at a, a + b, …, as indicated at the top of the panel. The inset depicts a kagomé net with thus determined local easy axes shown by the broken lines.

## IV. DISCUSSION

Summarizing our experimental results, CdK is a spin-1/2 Heisenberg KAFM with J = 45 K in the undistorted kagomé net and exhibits a canted antiferromagnetic order accompanied by a weak ferromagnetic moment only within the kagomé plane below $T_N$ = 4 K. From the torque measurements, the easy axes are determined as <100>. Now we discuss the possible magnetic structure of CdK.

### A. Dzyaloshinskii–Moriya interaction

Taking account of DM interactions, which should exist in the absence of inversion center at the middle of the bond between nearby Cu sites and must cause a spin canting as often observed in other KAFMs,[34] we consider the spin Hamiltonian given by

$$H = \sum_{ij}(J\, S_i \cdot S_j + D_{ij} \cdot S_i \times S_j), \quad (3)$$

where $D_{ij}$ is the DM vector between two sites i and j on the triangle; the convention is taken same as for the vector chirality in eq. (1). In CdK, since there is a mirror plane perpendicular to the bond, the DM vector should be confined in the mirror plane as depicted in Fig. 2(b); $D_z$ and $D_p$ are set to represent the out-of-plane and in-plane components, respectively; following the same convention used for the vector chirality, positive $D_z$ means a DM vector pointing upward from the paper in Fig. 2(b).[35]

The effects of the DM interaction on the classical KAFM have been theoretically studied, and a phase diagram of Fig. 8 is obtained by Monte Carlo simulations.[36,37] Upon switching the DM interaction, all the spins are forced to lie in the kagomé plane so that such coplanar spin structures as shown in Fig. 1 are selected; $D_z$ acts like an easy-plane anisotropy. Moreover, the **q** = 0 structures become more stable than the SVC structure and appear at $T \sim |D|$.[36] Importantly, the sign of $D_z$ selects the PVC or NVC, because it is explicitly coupled to the vector chirality in eq. (1): PVC is favored for negative $D_z$, while NVC for positive $D_z$; the $D_z$ term is cancelled out for SVC. On the other hand, $D_p$ tends to favor PVC so that large $|D_p|$ selects PVC even for positive $D_z$,[37] which causes the phase boundary bent in Fig. 8. However, because $|D_p|$ and $|D_z|$ are often smaller than J in real materials, one concludes that the sign of $D_z$ dominantly selects the spin structure.

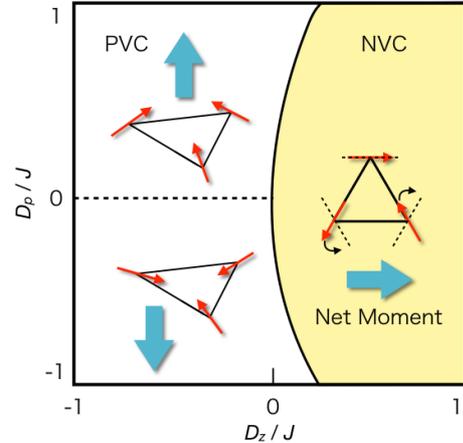

FIG. 8. Phase diagram for the ground state of the classical kagomé antiferromagnet with the nearest-neighbor antiferromagnetic interaction J and DM interaction.[36,37] Negative and positive $D_z$s tend to stabilize the PVC and NVC spin structures, respectively. In PVC, every spin is canted upward and downward by positive and negative $D_p$s, respectively, keeping the threefold rotational symmetry. This causes an out-of-plane net magnetization shown by the thick arrow. In NVC, spins are mostly within the plane and the threefold rotational symmetry is broken. The dashed lines on the triangle represent <100> easy axes. When one of the three spins on the triangle is pinned along the easy axis, the other two spins are not, which can generate a small tilting and cause an in-plane net magnetization.

In PVC, finite $D_p$ renders all spins to cant from the plane to an umbrella structure and leads to a spontaneous magnetization parallel to the c axis.[36] This is not the case of CdK but of the Cr jarosite.[31] Note that $D_p$ breaks the in-plane rotational symmetry and may induce in-plane anisotropy along <1−10>, which is different from the observed anisotropy along <100> for CdK.

Considering only DM interaction as a source of anisotropy, the NVC structure preserves the degree of freedom of global spin rotation in the plane. Nevertheless, it can be broken by a higher-order anisotropy term, and an in-plane net moment can occur. For example, when there is a <100> easy-axis anisotropy, a net magnetization is produced as illustrated in Fig. 8: in every triangle, only one moment can be parallel to the local easy axis, but the other two cannot, which may cause a canting that generates an uncompensated moment in the plane. A candidate for this higher-order anisotropy term is an anisotropic symmetric exchange interaction, which is usually much smaller than DM interactions.[38]

Provided that the magnetic moment is 1 $\mu_B$ / Cu, the spin rotation angle necessary to produce the observed weak ferromagnetic moment of 0.008 $\mu_B$ / Cu is calculated to be as small as 0.6°. This tiny value suggests that the higher order anisotropy energy is much smaller than $J$ which favors a 120° structure. This is also the reason for the indiscernible hysteresis in the magnetization curves of Fig. 6.

The above discussion is valid for classical KAFMs. However, as Cepas and coworkers show, a quantum KAFM takes a $\mathbf{q} = 0$ order for $|D_z|/J > 0.1$.[10] The magnitude of $D$ for CdK is roughly estimated by $(\Delta g/g)J \sim 0.1J$,[38] suggesting that CdK exists close to the boundary and probably in the ordered side. This value is also consistent with $T_N \sim |D| \sim 4$ K. Once a $\mathbf{q} = 0$ order sets in, the above classical arguments must be applied; quantum effects manifest themselves only in the contraction of a magnetic moment. Therefore, it is plausible to assume that the magnetic structure of CdK is NVC. Note that the presence of sizable DM interaction and the in-plane spontaneous magnetization strictly exclude SVC and PVC structures, and the observed easy axis of <100> is compatible with the NVC. Moreover, other coplanar or noncoplanar complex spin structures must not give such a weak ferromagnetism with distinct anisotropy.

The origin of the NVC order in CdK is apparently ascribed to the positive $D_z$, which happens to occur by the microscopic details of the compound.[36] On one hand, a similar NVC order has been established in the three-dimensional, metallic compounds, $Mn_3Sn$ and $Mn_3Ge$, which contain a stack of "breathing" kagomé nets of Mn moments of ~3 $\mu_B$.[39] Recently, they attract much attention because of the large anomalous Hall effects even in the antiferromagnets.[9,40,41] Probably, the better two-dimensionality and simpler magnetic interactions make CdK an optimized prototype of the NVC order

**B. Representation analysis**

To support the above discussion particularly on the direction of the weak ferromagnetic moment, a representation analysis has been carried out using the program SARA$h$ developed by Wills for representational analysis;[42] these results are basically same as the previous analysis for the jarosites.[43] The following implications are obtained:

1. twelve symmetry elements in the space group of $P-3m1$ make the propagation vector $\mathbf{k} = (0, 0, 0)$ invariant. The magnetic representation of a crystallographic site at (1/2, 0, 0) can be decomposed into irreducible representations of $\Gamma_1 + 2\Gamma_3 + 3\Gamma_6^2$, which basis vectors are listed in Table 1. Nine possible spin arrangements are schematically shown in Fig. 9. The PVC orders with <100> and <1–10> anisotropies belong to $\Gamma_1$ ($\psi_1$) and $\Gamma_3$ ($\psi_2$), respectively, and the NVC order belongs to $\Gamma_6$ ($2\psi_7 + \psi_8$).

2. for PVC, only out-of-plane weak ferromagnetism is allowed for <1–10> anisotropy by combining $\psi_2$ with $\psi_3$ having a ferromagnetic spin arrangement along $c$ in $\Gamma_3$. In contrast, such a canting is not allowed for <100> anisotropy in $\Gamma_1$.

3. for NVC, in-plane weak ferromagnetism is allowed because there is an in-plane ferromagnetic configuration in $\Gamma_6$; $-\psi_4 + \psi_5$ or $-\psi_7 + \psi_8$. A spin canting along $c$ is also possible, but no net moment should appear as there is no basis vector with a net moment along $c$ in $\Gamma_6$; all the spins in $\psi_9$ cancel with each other.

4. six symmetry elements make the propagation vector $\mathbf{k} = (1/3, 1/3, 0)$ invariant in the subgroup $G_\mathbf{q}$ of $P-3m1$. If symmetry elements are restricted to ones in $G_\mathbf{q}$, the situation is same as for $\Gamma_6$ in $\mathbf{k} = (0, 0, 0)$. Thus, a weak in-plane ferromagnetic moment is also allowed in SVC.

According to the representation analysis, the appearance of weak ferromagnetic moments only in the plane is not consistent with the PVC order but with either the NVC or the SVC order. Although the possibility of an SVC order is not excluded, finite DM interactions may prefer a NVC order in CdK, as mentioned above.

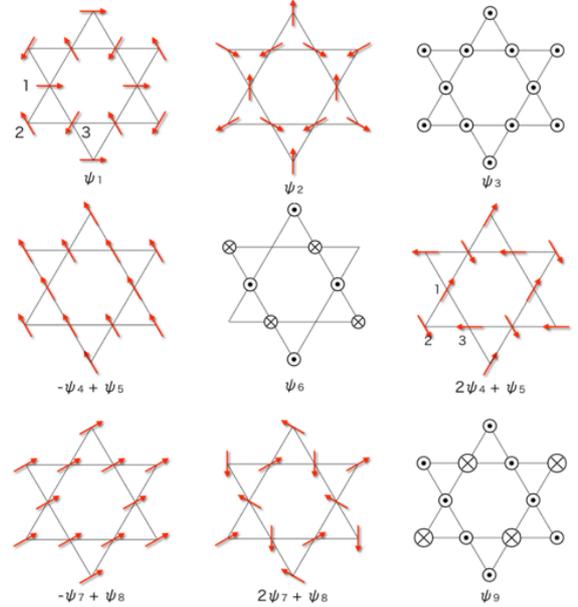

FIG. 9. Schematic representation of basis vectors for the space group $P-3m1$ with $\mathbf{k} = (0, 0, 0)$. A spin on each lattice point is represented by an arrow when it is confined in the plane, and by a circle when it is perpendicular to the plane. In $\psi_9$ one large downward spin and two small upward spins in the triangle cancel with each other.

| IR | BV | Atom | BV components | | |
|---|---|---|---|---|---|
| | | | $m_a$ | $m_b$ | $m_c$ |
| $\Gamma_1$ | $\psi_1$ | 1 | 4 | 0 | 0 |
| | | 2 | 0 | 4 | 0 |
| | | 3 | –4 | –4 | 0 |
| $\Gamma_3$ | $\psi_2$ | 1 | 2 | 3 | 0 |
| | | 2 | –4 | –2 | 0 |
| | | 3 | 2 | –2 | 0 |

| IR | BV | 1 | $m_a$ | $m_b$ | $m_c$ |
|---|---|---|---|---|---|
| | $\psi_3$ | 1 | 0 | 0 | 4 |
| | | 2 | 0 | 0 | 4 |
| | | 3 | 0 | 0 | 4 |
| | $\psi_4$ | 1 | 1 | 0 | 0 |
| | | 2 | 0 | –2 | 0 |
| | | 3 | –1 | –1 | 0 |
| | $\psi_5$ | 1 | 1 | 3 | 0 |
| | | 2 | 0 | 1 | 0 |
| | | 3 | –1 | 2 | 0 |
| $\Gamma_6$ | $\psi_6$ | 1 | 0 | 0 | 3 |
| | | 2 | 0 | 0 | 0 |
| | | 3 | 0 | 0 | –3 |
| | $\psi_7$ | 1 | –$\sqrt{3}$ | 0 | 0 |
| | | 2 | 0 | 0 | 0 |
| | | 3 | –$\sqrt{3}$ | –$\sqrt{3}$ | 0 |
| | $\psi_8$ | 1 | $\sqrt{3}$ | $\sqrt{3}$ | 0 |
| | | 2 | 2$\sqrt{3}$ | $\sqrt{3}$ | 0 |
| | | 3 | $\sqrt{3}$ | 0 | 0 |
| | $\psi_9$ | 1 | 0 | 0 | $\sqrt{3}$ |
| | | 2 | 0 | 0 | –2$\sqrt{3}$ |
| | | 3 | 0 | 0 | $\sqrt{3}$ |

Table. 1. Irreducible representations (IRs) and basis vectors (BVs) for the space group $P$–3$m$1 with **k** = (0, 0, 0). The decomposition of the magnetic structure representation for the 3$e$ site (1/2, 0, 0) is $\Gamma_{mag} = \Gamma_1 + \Gamma_3 + 3\Gamma_6^2$. The atoms in the nonprimitive basis are defined as 1 (1/2, 0, 0), 2 (0, 1/2, 0), and 3 (1/2, 1/2, 0). The BV components along the crystallographic axes are shown by $m_i$ ($i = a, b, c$).

### C. Critical phenomenon

Finally, we point out one intriguing feature of the NVC order with <100> anisotropy as realized in CdK and its relation to the broad peak in the heat capacity at the transition. The canted NVC structure or any magnetic structure with spin canting that causes an in-plane weak ferromagnetic moment spontaneously breaks $Z_6$ anisotropy. In three dimension, numerical calculations on the XY model with $Z_6$ anisotropy suggest that the critical phenomena belong to the 3D XY universality class.[44-46] Interestingly, this type of critical behavior may be closely related to the "deconfined" quantum criticality.[47-49] Thus, it would be interesting to investigate the critical behavior of the transition to NVC that breaks the $Z_6$ symmetry in CdK. One implication to experiments is that the transition may not be accompanied by a divergence in heat capacity. The observed broad peak in the heat capacity of CdK is possibly intrinsic and may be related to the unique characteristics of the transition to NVC.

### V. CONCLUSION

We have successfully grown single crystals of the kagomé antiferromagnet CdCu$_3$(OH)$_6$(NO$_3$)$_2$•H$_2$O with $J$ = 45 K, and suggest a NVC spin order below $T_N$ ~ 4 K. The origin of the NVC order is considered to be a DM interaction with a positive $z$ component. We believe that there are novel phenomena for this NVC order to be searched for in the future work.

### ACKNOWLEDGEMENTS

We are grateful to Yoshihiko Okamoto, Hajime Ishikawa, Kazuhiro Nawa, Nic Shannon, and Naoki Kawashima for helpful discussion. We also thank B. Canal, G. J. Nilsen and S. Hayashida for valuable comments. RO is supported by the Materials Education Program for the Future Leaders in Research, Industry, and Technology (MERIT) given by the Ministry of Education, Culture, Sports, Science and Technology of Japan (MEXT). This work was partially supported by KAKENHI (Grant Number 15K17701) and the Core-to-Core Program for Advanced Research Networks given by the Japan Society for the Promotion of Science (JSPS). It was also supported by the Strategic Programs for Innovative Research (SPIRE), MEXT and the Computational Materials Science Initiative (CMSI), Japan.